\documentclass{amsbook}
\usepackage{amsmath}

\theoremstyle{plain}

\newtheorem{definition}{Definition}
\newtheorem{example}{Example}

\newtheorem{notation}{Notation}

\newtheorem{remark}{Remark}

\newtheorem{theorem}{Theorem}
\numberwithin{equation}{section}
\input{tcilatex}

\begin{document}
\frontmatter
\title[ \ ]{The intersection and the union of the asynchronous systems}
\author{Serban E. Vlad}
\address{Str. Zimbrului, Nr. 3, Bl. PB 68, Ap. 11, 410430, Oradea, Romania}
\email{serban\_e\_vlad@yahoo.com}
\urladdr{http://www.geocities.com/serban\_e\_vlad}
\maketitle
\tableofcontents

\mainmatter

\begin{center}
\textbf{Abstract}
\end{center}

The asynchronous systems $f$ are the models of the asynchronous circuits
from digital electrical engineering. They are multi-valued functions that
associate to each input $u:\mathbf{R}\rightarrow \{0,1\}^{m}$ a set of
states $x\in f(u),$ where $x:\mathbf{R}\rightarrow \{0,1\}^{n}.$ The
intersection of the systems allows adding supplementary conditions in
modeling and the union of the systems allows considering the validity of one
of two systems in modeling, for example when testing the asynchronous
circuits and the circuit is supposed to be 'good' or 'bad'. The purpose of
the paper is that of analyzing the intersection and the union against the
initial/final states, initial/final time, initial/final state functions,
subsystems, dual systems, inverse systems, Cartesian product of systems,
parallel connection and serial connection of systems.

\begin{center}
\textbf{Rezumat}
\end{center}

Sistemele asincrone $f$ sunt modelele circuitelor asincrone din electronica
digitala. Ele sunt functii multivoce care asociaza fiecarei intrari $u:%
\mathbf{R}\rightarrow \{0,1\}^{m}$ o multime de stari $x\in f(u),$ unde $x:%
\mathbf{R}\rightarrow \{0,1\}^{n}.$ Intersectia sistemelor permite adaugarea
de conditii suplimentare in modelare si reuniunea sistemelor permite
considerarea validitatii unui sistem din doua in modelare, de exemplu cand
se testeaza circuitele asincrone si circuitul se presupune a fi 'bun' sau
'rau'. Scopul lucrarii e acela de a analiza intersectia si reuniunea
raportate la starile initiale/finale, timpul initial/final, functiile stare
initiala/finala, subsisteme, sisteme duale, sisteme inverse, produs
cartezian al sistemelor, legare in paralel si legare in serie a sistemelor.



\section{Preliminary definitions}

\begin{definition}
The set $\mathbf{B}=\{0,1\}$ endowed with the laws: the complement '$%
\overline{\quad }$', the union '$\cup $', the intersection '$\cdot $', the
modulo 2 sum '$\oplus $' etc is called the binary Boole algebra.
\end{definition}

\begin{definition}
\label{Def2}We denote by $\mathbf{R}$ the set of the real numbers. The
initial value $x(-\infty +0)\in \mathbf{B}$ and the final value $x(\infty
-0)\in \mathbf{B}$ of the function $x:\mathbf{R}\rightarrow \mathbf{B}$ are
defined by%
\begin{equation*}
\exists t_{0}\in \mathbf{R},\forall t<t_{0},x(t)=x(-\infty +0)
\end{equation*}%
\begin{equation*}
\exists t_{f}\in \mathbf{R},\forall t>t_{f},x(t)=x(\infty -0)
\end{equation*}%
The definition and the notations are similar for the $\mathbf{R}\rightarrow 
\mathbf{B}^{n}$ functions, $n\geq 1.$
\end{definition}

\begin{definition}
The characteristic function $\chi _{A}:\mathbf{R}\rightarrow \mathbf{B}$ of
the set $A\subset \mathbf{R}$ is defined by%
\begin{equation*}
\forall t\in \mathbf{R},\chi _{A}(t)=\left\{ 
\begin{array}{c}
1,if\;t\in A \\ 
0,if\;t\notin A%
\end{array}%
\right.
\end{equation*}
\end{definition}

\begin{definition}
The set $S^{(n)}$ of the $n-$signals consists by definition in the functions 
$x:\mathbf{R}\rightarrow \mathbf{B}^{n}$ of the form%
\begin{equation*}
x(t)=x(-\infty +0)\cdot \chi _{(-\infty ,t_{0})}(t)\oplus x(t_{0})\cdot \chi
_{\lbrack t_{0},t_{1})}(t)\oplus x(t_{1})\cdot \chi _{\lbrack
t_{1},t_{2})}(t)\oplus ...
\end{equation*}%
where $x(-\infty +0)\in \mathbf{B}^{n},$ $t_{0}<t_{1}<t_{2}<...$ is some
strictly increasing unbounded sequence of real numbers and the laws '$\cdot $%
', '$\oplus $' are induced by those from $\mathbf{B}$.
\end{definition}

\begin{notation}
For an arbitrary set $H$, we use the notation%
\begin{equation*}
P^{\ast }(H)=\{H^{\prime }|H^{\prime }\subset H,H^{\prime }\neq \emptyset \}
\end{equation*}
\end{notation}

\begin{definition}
\label{Def5}The functions $f:U\rightarrow P^{\ast }(S^{(n)}),U\in P^{\ast
}(S^{(m)})$ are called (asynchronous) systems. Any $u\in U$ is called
(admissible) input of $f$ and the functions $x\in f(u)$ are the (possible)
states of $f$.
\end{definition}

\begin{remark}
In the paper $t\in \mathbf{R}$ represents time. The $n-$signals model the
tensions in digital electrical engineering and the asynchronous systems are
the models of the asynchronous circuits. They represent multi-valued
associations between a cause $u$ and a set $f(u)$ of effects because of the
uncertainties that occur in modeling.

Definition \ref{Def5} represents the definition of the systems given under
the explicit form. In previous works (such as \cite{bib40}) we used
equations and inequalities for defining systems under the implicit form.
\end{remark}

\begin{definition}
\label{Def44}We have the systems $f:U\rightarrow P^{\ast }(S^{(n)}),$ $%
g:V\rightarrow P^{\ast }(S^{(n)})$ with $U,V\in P^{\ast }(S^{(m)})$. If $%
\exists u\in U\cap V,f(u)\cap g(u)\neq \emptyset $, the system $f\cap
g:W\rightarrow P^{\ast }(S^{(n)})$ defined by%
\begin{equation}
W=\{u|u\in U\cap V,f(u)\cap g(u)\neq \emptyset \}  \label{ir_1}
\end{equation}%
\begin{equation*}
\forall u\in W,(f\cap g)(u)=f(u)\cap g(u)
\end{equation*}%
is called the intersection of $f$ and $g$.
\end{definition}

\begin{remark}
The intersection of the systems represents the gain of information (of
precision) in the modeling of a circuit that results by considering the
validity of two (compatible!) models at the same time.

We have the special case when $V=S^{(m)}$ and the system $g$ is constant
(such systems are called autonomous): $\forall u\in S^{(m)},$ $g(u)=X$ where 
$X\in P^{\ast }(S^{(n)})$. Then $f\cap X:W\rightarrow P^{\ast }(S^{(n)})$ is
the system given by%
\begin{equation*}
W=\{u|u\in U,f(u)\cap X\neq \emptyset \}
\end{equation*}%
\begin{equation*}
\forall u\in W,(f\cap X)(u)=f(u)\cap X
\end{equation*}%
We interpret $f\cap X$ in the next manner. When $f$ models a circuit, $f\cap
X$ represents a gain of information resulting by the statement of a request
that does not depend on $u$.
\end{remark}

\begin{example}
We give some possibilities of choosing in the intersection $f\cap g$ the
constant system $g=X:$

i) the initial value of the states is null

ii) the coordinates $x_{1},...,x_{n}$ of the states are monotonous relative
to the order $0<1$ (this allows defining the so called hazard-freedom of the
systems)

iii) at each time instant, at least one coordinate of the state should be 1:%
\begin{equation*}
X=\{x|x\in S^{(n)},\forall t\in \mathbf{R},x_{1}(t)\cup ...\cup x_{n}(t)=1\}
\end{equation*}

iv) the state can switch\footnote{%
the left limit $x(t-0)$ of $x(t)$ that occurs in some examples is defined
like this:%
\begin{equation*}
\forall t\in \mathbf{R},\exists \varepsilon >0,\forall \xi \in
(t-\varepsilon ,t),x(\xi )=x(t-0)
\end{equation*}%
$x$ switches if $x(t-0)\neq x(t)$, i.e. if it has a (left) discontinuity}
with at most one coordinate at a time (a special case when the so called
technical condition of good running of the systems is satisfied):%
\begin{equation*}
X=\{x|x\in S^{(n)},\forall t\in \mathbf{R},x(t-0)\neq x(t)\Longrightarrow
\exists !i\in \{1,...,n\},x_{i}(t-0)\neq x_{i}(t)\}
\end{equation*}

v) $X$ represents a 'stuck at 1 fault':%
\begin{equation*}
\exists i\in \{1,...,n\},X=\{x|x\in S^{(n)},\forall t\in \mathbf{R}%
,x_{i}(t)=1\}
\end{equation*}%
This last choice of $X$ is interesting in designing systems for testability,
respectively in designing redundant systems.

vi) $X$ consists in all $x\in S^{(n)}$ satisfying the next 'absolute
inertia' property: $\delta _{r}>0,\delta _{f}>0$ are given so that $\forall
i\in \{1,...,n\},\forall t\in \mathbf{R},$%
\begin{equation*}
\overline{x_{i}(t-0)}\cdot x_{i}(t)\leq \underset{\xi \in \lbrack t,t+\delta
_{r}]}{\bigcap }x_{i}(\xi )
\end{equation*}%
\begin{equation*}
x_{i}(t-0)\cdot \overline{x_{i}(t)}\leq \underset{\xi \in \lbrack t,t+\delta
_{f}]}{\bigcap }\overline{x_{i}(\xi )}
\end{equation*}%
The interpretation of these inequalities is the following: if $x_{i}$
switches from 0 to 1, then it remains 1 for more than $\delta _{r}$ time
units and if $x_{i}$ switches from 1 to 0 then it remains 0 for more than $%
\delta _{f}$ time units.
\end{example}

\begin{example}
We show a possibility of choosing in the intersection $f\cap g$, $g$
non-constant. The Boolean function $F:\mathbf{B}^{m}\rightarrow \mathbf{B}%
^{n}$ is given and $f$ is the arbitrary model of a circuit that computes $F$%
. $V=S^{(m)}$ and the parameters $\delta _{r}>0,\delta _{f}>0$ exist so that%
\begin{equation*}
\forall u\in S^{(m)},g(u)=\{x|x\in S^{(n)},\forall i\in \{1,...,n\},\forall
t\in \mathbf{R},
\end{equation*}%
\begin{equation*}
\overline{x_{i}(t-0)}\cdot x_{i}(t)\leq \underset{\xi \in \lbrack t-\delta
_{r},t)}{\bigcap }F_{i}(u(\xi ))
\end{equation*}%
\begin{equation*}
x_{i}(t-0)\cdot \overline{x_{i}(t)}\leq \underset{\xi \in \lbrack t-\delta
_{f},t)}{\bigcap }\overline{F_{i}(u(\xi ))}\}
\end{equation*}%
meaning that $g(u)$ contains all $x$ with the property that, on all the
coordinates $i$ and at all the time instants $t$:

- $x_{i}$ switches from $0$ to $1$ only if $F_{i}(u(\cdot ))$ was $1$ for at
least $\delta _{r}$ time units

- $x_{i}$ switches from $1$ to $0$ only if $F_{i}(u(\cdot ))$ was $0$ for at
least $\delta _{f}$ time units.
\end{example}

\begin{definition}
\label{Def45}The union of the systems $f:U\rightarrow P^{\ast }(S^{(n)})$
and $g:V\rightarrow P^{\ast }(S^{(n)}),U,V\in P^{\ast }(S^{(m)})$ is the
system $f\cup g:U\cup V\rightarrow P^{\ast }(S^{(n)})$ that is defined by%
\begin{equation*}
\forall u\in U\cup V,(f\cup g)(u)=\left\{ 
\begin{array}{c}
f(u),if\;u\in U\setminus V \\ 
g(u),if\;u\in V\setminus U \\ 
f(u)\cup g(u),if\;u\in U\cap V%
\end{array}%
\right.
\end{equation*}%
If $U\cap V=\emptyset ,$ then $f\cup g$ is called the disjoint union of $f$
and $g$.
\end{definition}

\begin{remark}
The union of the systems is the dual concept to that of intersection
representing the loss of information (of precision) in modeling that results
in general by considering the validity of one of two models of the same
circuit. The disjoint union means no loss of information however.

Another possibility is that in Definition \ref{Def45} $f,g$ model two
different circuits, see Example \ref{Exa3}.

We have the special case when in the union $f\cup g$ the system $g$ is
constant under the form $V=S^{(m)},$ $g:S^{(m)}\rightarrow P^{\ast
}(S^{(n)}),$ $\forall u\in S^{(m)},$ $g(u)=X,$ with $X\subset S^{(n)}.$ Then 
$f\cup X:S^{(m)}\rightarrow P^{\ast }(S^{(n)})$ is defined by:%
\begin{equation*}
\forall u\in S^{(m)},(f\cup X)(u)=\left\{ 
\begin{array}{c}
X,if\;u\in S^{(m)}\setminus U \\ 
f(u)\cup X,if\;u\in U%
\end{array}%
\right. 
\end{equation*}%
The interpretation of $f\cup X$ is the next one: when $f$ is the model of an
asynchronous circuit, $X$ represents perturbations that are independent on $u
$.
\end{remark}

\begin{example}
\label{Exa3}In the union $f\cup g$ we presume that $U\cap V\neq \emptyset $
and $f$, $g$ model two different circuits, the first considered 'good,
without errors' and the second 'bad, with a certain error'. The testing
problem consists in finding an input $u\in U\cap V$ so that $f(u)\cap
g(u)=\emptyset ;$ after its application to $f\cup g$ and the measurement of
a state $x\in (f\cup g)(u),$ we can say if $x\in f(u)$ and the tested
circuit is 'good' or perhaps $x\in g(u)$ and the tested circuit is 'bad'.
\end{example}

\section{Initial states and final states}

\begin{remark}
In the next properties of the system $f$:%
\begin{equation}
\forall u\in U,\forall x\in f(u),\exists \mu \in \mathbf{B}^{n},\exists
t_{0}\in \mathbf{R},\forall t<t_{0},x(t)=\mu  \label{isfs1}
\end{equation}%
\begin{equation}
\forall u\in U,\exists \mu \in \mathbf{B}^{n},\forall x\in f(u),\exists
t_{0}\in \mathbf{R},\forall t<t_{0},x(t)=\mu  \label{isfs2}
\end{equation}%
\begin{equation}
\exists \mu \in \mathbf{B}^{n},\forall u\in U,\forall x\in f(u),\exists
t_{0}\in \mathbf{R},\forall t<t_{0},x(t)=\mu  \label{isfs3}
\end{equation}%
\begin{equation}
\forall u\in U,\forall x\in f(u),\exists \mu \in \mathbf{B}^{n},\exists
t_{f}\in \mathbf{R},\forall t\geq t_{f},x(t)=\mu  \label{isfs4}
\end{equation}%
\begin{equation}
\forall u\in U,\exists \mu \in \mathbf{B}^{n},\forall x\in f(u),\exists
t_{f}\in \mathbf{R},\forall t\geq t_{f},x(t)=\mu  \label{isfs5}
\end{equation}%
\begin{equation}
\exists \mu \in \mathbf{B}^{n},\forall u\in U,\forall x\in f(u),\exists
t_{f}\in \mathbf{R},\forall t\geq t_{f},x(t)=\mu  \label{isfs6}
\end{equation}%
we have replaced $t>t_{f}$ from Definition \ref{Def2} with $t\geq t_{f}$ and
on the other hand (\ref{isfs1}) is always true due to the way that the $n-$%
signals were defined. We remark the truth of the implications%
\begin{equation*}
(\ref{isfs3})\Longrightarrow (\ref{isfs2})\Longrightarrow (\ref{isfs1})
\end{equation*}%
\begin{equation*}
(\ref{isfs6})\Longrightarrow (\ref{isfs5})\Longrightarrow (\ref{isfs4})
\end{equation*}
\end{remark}

\begin{definition}
\label{Def21}Because $f$ satisfies (\ref{isfs1}), we use to say that it has
initial states. The vectors $\mu $ are called (the) initial states (of $f$),
or (the) initial values of the states (of $f$).
\end{definition}

\begin{definition}
\label{Def21_}We presume that $f$ satisfies (\ref{isfs2}). We say in this
situation that it has race-free initial states and the initial states $\mu $
are called race-free themselves.
\end{definition}

\begin{definition}
\label{Def22}When $f$ satisfies (\ref{isfs3}), we use to say that it has a
(constant) initial state $\mu $. We say in this case that $f$ is initialized
and that $\mu $ is its (constant) initial state.
\end{definition}

\begin{definition}
\label{Def23}If $f$ satisfies (\ref{isfs4}), it is called absolutely stable
and we also say that it has final states. The vectors $\mu $ have in this
case the name of final states (of $f$), or of final values of the states (of 
$f$).
\end{definition}

\begin{definition}
\label{Def24}If $f$ fulfills the property (\ref{isfs5}), it is called
absolutely race-free stable and we also say that it has race-free final
states. The final states $\mu $ are called in this case race-free.
\end{definition}

\begin{definition}
\label{Def25}We presume that the system $f$ satisfies (\ref{isfs6}). Then it
is called absolutely constantly stable or equivalently we say that it has a
(constant) final state. The vector $\mu $ is called in this situation
(constant) final state.
\end{definition}

\begin{theorem}
Let $f:U\rightarrow P^{\ast }(S^{(n)})$ and $g:V\rightarrow P^{\ast
}(S^{(n)})$ be some systems$,$ $U,V\in P^{\ast }(S^{(m)})$. If $\exists u\in
U\cap V,f(u)\cap g(u)\neq \emptyset $ and $f$ has race-free initial states
(constant initial state), then $f\cap g$ has race-free initial states
(constant initial state).
\end{theorem}

\begin{proof}
If one of the previous properties is true for the states in $f(u)$, then it
is true for the states in the subset $f(u)\cap g(u)\subset f(u)$ also, $u\in
U$.
\end{proof}

\begin{theorem}
\label{The78}If $f$ has final states (race-free final states, constant final
state) and $f\cap g$ exists, then $f\cap g$ has final states (race-free
final states, constant final state).
\end{theorem}

\begin{theorem}
\label{The88}a) If $f,g$ have race-free initial states and $\forall u\in
U\cap V,f(u)\cap g(u)\neq \emptyset $ then $f\cup g$ has race-free initial
states.

b) If $f,g$ have constant initial states and $\underset{u\in U}{\bigcup }%
f(u)\cap \underset{u\in V}{\bigcup }g(u)\neq \emptyset $ then $f\cup g$ has
constant initial states.
\end{theorem}

\begin{proof}
a) The hypothesis states the truth of the next properties%
\begin{equation*}
\forall u\in U,\exists \mu \in \mathbf{B}^{n},\forall x\in f(u),\exists
t_{0}\in \mathbf{R},\forall t<t_{0},x(t)=\mu
\end{equation*}%
\begin{equation*}
\forall u\in V,\exists \mu \in \mathbf{B}^{n},\forall x\in g(u),\exists
t_{0}\in \mathbf{R},\forall t<t_{0},x(t)=\mu
\end{equation*}%
\begin{equation}
\forall u\in U\cap V,f(u)\cap g(u)\neq \emptyset  \label{isfs7}
\end{equation}%
If $(U\setminus V)\cup (V\setminus U)\neq \emptyset ,$ then $\forall u\in
(U\setminus V)\cup (V\setminus U)$ the statement is true because it states
separately for $f$ and $g$ that they have race-free initial states. And if $%
U\cap V\neq \emptyset ,$ then $\forall u\in U\cap V,\forall x\in f(u)\cup
g(u),$ the initial value $\mu =x(-\infty +0)$ depends on $u$ only, not also
on the fact that $x\in f(u)$ or $x\in g(u)$ due to (\ref{isfs7})$.$ We have
that%
\begin{equation*}
\forall u\in U\cup V,\exists \mu \in \mathbf{B}^{n},\forall x\in (f\cup
g)(u),\exists t_{0}\in \mathbf{R},\forall t<t_{0},x(t)=\mu
\end{equation*}%
is true.

b) Because $\underset{u\in U}{\bigcup }f(u)\cap \underset{u\in V}{\bigcup }%
g(u)\neq \emptyset ,$ in the statements%
\begin{equation*}
\exists \mu \in \mathbf{B}^{n},\forall u\in U,\forall x\in f(u),\exists
t_{0}\in \mathbf{R},\forall t<t_{0},x(t)=\mu
\end{equation*}%
\begin{equation*}
\exists \mu ^{\prime }\in \mathbf{B}^{n},\forall u\in V,\forall x\in
g(u),\exists t_{0}\in \mathbf{R},\forall t<t_{0},x(t)=\mu ^{\prime }
\end{equation*}%
the two constants $\mu $ and $\mu ^{\prime }$, whose existence is unique,
coincide.
\end{proof}

\begin{theorem}
\label{The89}a) If $f,g$ have final states, then $f\cup g$ has final states.

b) If $f,g$ have race-free final states and $\forall u\in U\cap V,f(u)\cap
g(u)\neq \emptyset $ then $f\cup g$ has race-free final states.

c) If $f,g$ have constant final states and $\underset{u\in U}{\bigcup }%
f(u)\cap \underset{u\in V}{\bigcup }g(u)\neq \emptyset $ then $f\cup g$ has
constant final states.
\end{theorem}

\section{Initial time and final time}

\begin{notation}
The set of the $n-$signals with final values is denoted by $S_{c}^{(n)}.$ It
consists in the functions $x:\mathbf{R}\rightarrow \mathbf{B}^{n}$ of the
form%
\begin{equation*}
x(t)=x(-\infty +0)\cdot \chi _{(-\infty ,t_{0})}(t)\oplus x(t_{0})\cdot \chi
_{\lbrack t_{0},t_{1})}(t)\oplus x(t_{1})\cdot \chi _{\lbrack
t_{1},t_{2})}(t)\oplus ...
\end{equation*}%
\begin{equation*}
...\oplus x(t_{k})\cdot \chi _{\lbrack t_{k},t_{k+1})}(t)\oplus x(\infty
-0)\cdot \chi _{\lbrack t_{k+1},\infty )}(t)
\end{equation*}%
where $x(-\infty +0),x(\infty -0)\in \mathbf{B}^{n}$ and $%
t_{0}<t_{1}<...<t_{k}<t_{k+1}$ is a finite family of real numbers, $k\geq 0.$
\end{notation}

\begin{remark}
We state the next properties on the asynchronous system $f:U\rightarrow
P^{\ast }(S^{(n)}),U\in P^{\ast }(S^{(m)})$:%
\begin{equation}
\forall u\in U,\forall x\in f(u),\exists \mu \in \mathbf{B}^{n},\exists
t_{0}\in \mathbf{R},\forall t<t_{0},x(t)=\mu  \label{itft1}
\end{equation}%
\begin{equation}
\forall u\in U,\exists t_{0}\in \mathbf{R},\forall x\in f(u),\exists \mu \in 
\mathbf{B}^{n},\forall t<t_{0},x(t)=\mu  \label{itft2}
\end{equation}%
\begin{equation}
\exists t_{0}\in \mathbf{R},\forall u\in U,\forall x\in f(u),\exists \mu \in 
\mathbf{B}^{n},\forall t<t_{0},x(t)=\mu  \label{itft3}
\end{equation}%
\begin{equation}
\forall u\in U,\forall x\in f(u)\cap S_{c}^{(n)},\exists \mu \in \mathbf{B}%
^{n},\exists t_{f}\in \mathbf{R},\forall t\geq t_{f},x(t)=\mu  \label{itft4}
\end{equation}%
\begin{equation}
\forall u\in U,\exists t_{f}\in \mathbf{R},\forall x\in f(u)\cap
S_{c}^{(n)},\exists \mu \in \mathbf{B}^{n},\forall t\geq t_{f},x(t)=\mu
\label{itft5}
\end{equation}%
\begin{equation}
\exists t_{f}\in \mathbf{R},\forall u\in U,\forall x\in f(u)\cap
S_{c}^{(n)},\exists \mu \in \mathbf{B}^{n},\forall t\geq t_{f},x(t)=\mu
\label{itft6}
\end{equation}%
The properties (\ref{itft1}) and (\ref{itft4}) are fulfilled by all the
systems and the next implications hold:%
\begin{equation*}
(\ref{itft3})\Longrightarrow (\ref{itft2})\Longrightarrow (\ref{itft1})
\end{equation*}%
\begin{equation*}
(\ref{itft6})\Longrightarrow (\ref{itft5})\Longrightarrow (\ref{itft4})
\end{equation*}
\end{remark}

\begin{definition}
\label{Def26}The fact that $f$ satisfies (\ref{itft1}) is expressed
sometimes by saying that it has unbounded initial time and any $t_{0}$
satisfying this property is called unbounded initial time (instant).
\end{definition}

\begin{definition}
\label{Def27}Let $f$ be a system that fulfills the property (\ref{itft2}).
We say that it has bounded initial time and any $t_{0}$ making this property
true is called bounded initial time (instant).
\end{definition}

\begin{definition}
\label{Def28}When $f$ satisfies (\ref{itft3}), we use to say that it has
fixed initial time and any $t_{0}$ fulfilling (\ref{itft3}) is called fixed
initial time (instant).
\end{definition}

\begin{definition}
\label{Def29}The fact that $f$ satisfies (\ref{itft4}) is expressed by
saying that it has unbounded final time and any $t_{f}$ satisfying this
property is called unbounded final time (instant).
\end{definition}

\begin{definition}
\label{Def30}If $f$ fulfills the property (\ref{itft5}), we say that it has
bounded final time. Any number $t_{f}$ satisfying (\ref{itft5}) is called
bounded final time (instant).
\end{definition}

\begin{definition}
\label{Def31}We presume that the system $f$ satisfies the property (\ref%
{itft6}). Then we say that it has fixed final time and any number $t_{f}$
satisfying (\ref{itft6}) is called fixed final time (instant).
\end{definition}

\begin{theorem}
If $f$ has bounded initial time (fixed initial time) and $f\cap g$ exists,
then $f\cap g$ has bounded initial time (fixed initial time).
\end{theorem}

\begin{proof}
Like previously, if one of the above properties is true for the states in $%
f(u),$ then it is true for the states in $f(u)\cap g(u)\subset f(u),u\in U$.
\end{proof}

\begin{theorem}
\label{The_80}If $f$ has bounded final time (fixed final time) and $f\cap g$
exists, then $f\cap g$ has bounded final time (fixed final time).
\end{theorem}

\begin{theorem}
If $f,g$ have bounded initial time (fixed initial time), then $f\cup g$ has
bounded initial time (fixed initial time).
\end{theorem}

\begin{proof}
We presume that $f,g$ have bounded initial time. If $(U\setminus V)\cup
(V\setminus U)\neq \emptyset ,$ then $\forall u\in (U\setminus V)\cup
(V\setminus U),$ $(f\cup g)(u)$ has the desired property, that refers to
exactly one of $f,g.$ We presume that $U\cap V\neq \emptyset $ and let $u\in
U\cap V$ be arbitrary. $t_{0}^{\prime },t_{0}^{"}\in \mathbf{R}$ exist,
depending on $u$, so that%
\begin{equation*}
\forall x\in f(u),\exists \mu \in \mathbf{B}^{n},\forall t<t_{0}^{\prime
},x(t)=\mu 
\end{equation*}%
\begin{equation*}
\forall x\in g(u),\exists \mu \in \mathbf{B}^{n},\forall
t<t_{0}^{"},x(t)=\mu 
\end{equation*}%
$t_{0}=\min \{t_{0}^{\prime },t_{0}^{"}\}$ satisfies%
\begin{equation*}
\forall x\in f(u)\cup g(u),\exists \mu \in \mathbf{B}^{n},\forall
t<t_{0},x(t)=\mu 
\end{equation*}
\end{proof}

\begin{theorem}
\label{The93}If $f,g$ have bounded final time (fixed final time), then $%
f\cup g$ has bounded final time (fixed final time).
\end{theorem}

\section{Initial state function and set of initial states. Final state
function and set of final states}

\begin{definition}
\label{Def32}Let $f:U\rightarrow P^{\ast }(S^{(n)}),U\in P^{\ast }(S^{(m)})$
be a system. The initial state function $\phi _{0}:U\rightarrow P^{\ast }(%
\mathbf{B}^{n})$ and the set of the initial states $\Theta _{0}\in P^{\ast }(%
\mathbf{B}^{n})$ of $f$ are defined by%
\begin{equation*}
\forall u\in U,\phi _{0}(u)=\{x(-\infty +0)|x\in f(u)\}
\end{equation*}%
\begin{equation*}
\Theta _{0}=\underset{u\in U}{\bigcup }\phi _{0}(u)
\end{equation*}
\end{definition}

\begin{definition}
\label{Def33}If $f$ has final states, i.e. if (\ref{isfs4}) is satisfied,
the final state function $\phi _{f}:U\rightarrow P^{\ast }(\mathbf{B}^{n})$
and the set of the final states $\Theta _{f}\in P^{\ast }(\mathbf{B}^{n})$
of $f$ are%
\begin{equation*}
\forall u\in U,\phi _{f}(u)=\{x(\infty -0)|x\in f(u)\}
\end{equation*}%
\begin{equation*}
\Theta _{f}=\underset{u\in U}{\bigcup }\phi _{f}(u)
\end{equation*}
\end{definition}

\begin{theorem}
\label{The80}For the systems $f,g$ we have $(\phi \cap \gamma
)_{0}:W\rightarrow P^{\ast }(\mathbf{B}^{n}),$%
\begin{equation*}
\forall u\in W,(\phi \cap \gamma )_{0}(u)=\phi _{0}(u)\cap \gamma _{0}(u)
\end{equation*}%
\begin{equation*}
(\Theta \cap \Gamma )_{0}=\underset{u\in W}{\bigcup }(\phi \cap \gamma
)_{0}(u)
\end{equation*}%
We have presumed that the domain $W$ of $f\cap g$ is non-empty and we have
denoted by $\phi _{0},\gamma _{0},(\phi \cap \gamma )_{0}$ the initial state
functions of $f,g,f\cap g$ and respectively by $(\Theta \cap \Gamma )_{0}$
the set of initial states of $f\cap g$.
\end{theorem}

\begin{proof}
We can write that $\forall u\in W,$%
\begin{equation*}
(\phi \cap \gamma )_{0}(u)=\{x(-\infty +0)|x\in (f\cap g)(u)\}=\{x(-\infty
+0)|x\in f(u)\cap g(u)\}=
\end{equation*}%
\begin{equation*}
=\{x(-\infty +0)|x\in f(u)\}\cap \{x(-\infty +0)|x\in g(u)\}=\phi
_{0}(u)\cap \gamma _{0}(u)
\end{equation*}
\end{proof}

\begin{theorem}
\label{The_82}If $f,g$ have final states, then we have $(\phi \cap \gamma
)_{f}:W\rightarrow P^{\ast }(\mathbf{B}^{n}),$%
\begin{equation*}
\forall u\in W,(\phi \cap \gamma )_{f}(u)=\phi _{f}(u)\cap \gamma _{f}(u)
\end{equation*}%
\begin{equation*}
(\Theta \cap \Gamma )_{f}=\underset{u\in W}{\bigcup }(\phi \cap \gamma
)_{f}(u)
\end{equation*}%
We have presumed that $W\neq \emptyset $ and the notations are obvious and
similar with those from the previous theorem.
\end{theorem}

\begin{theorem}
\label{The92}For the systems $f,g$ we have $(\phi \cup \gamma )_{0}:U\cup
V\rightarrow P^{\ast }(\mathbf{B}^{n}),$%
\begin{equation*}
\forall u\in U\cup V,(\phi \cup \gamma )_{0}(u)=\left\{ 
\begin{array}{c}
\phi _{0}(u),\;u\in U\setminus V \\ 
\gamma _{0}(u),\;u\in V\setminus U \\ 
\phi _{0}(u)\cup \gamma _{0}(u),\;u\in U\cap V%
\end{array}%
\right.
\end{equation*}%
\begin{equation*}
(\Theta \cup \Gamma )_{0}=\underset{u\in U\cup V}{\bigcup }(\phi \cup \gamma
)_{0}(u)
\end{equation*}%
We have denoted by $(\phi \cup \gamma )_{0}$ the initial state function of $%
f\cup g$ and respectively by $(\Theta \cup \Gamma )_{0}$ the set of initial
states of $f\cup g$.
\end{theorem}

\begin{proof}
Three possibilities exist for an arbitrary $u\in U\cup V:$ $u\in U\setminus
V,u\in V\setminus U$ and $u\in U\cap V.$ If for example $u\in U\setminus V,$
then:%
\begin{equation*}
(\phi \cup \gamma )_{0}(u)=\{x(-\infty +0)|x\in (f\cup g)(u)\}=\{x(-\infty
+0)|x\in f(u)\}=\phi _{0}(u)
\end{equation*}
\end{proof}

\begin{theorem}
\label{The95}We presume that $f,g$ have final states. We have $(\phi \cup
\gamma )_{f}:U\cup V\rightarrow P^{\ast }(\mathbf{B}^{n}),$%
\begin{equation*}
\forall u\in U\cup V,(\phi \cup \gamma )_{f}(u)=\left\{ 
\begin{array}{c}
\phi _{f}(u),\;u\in U\setminus V \\ 
\gamma _{f}(u),\;u\in V\setminus U \\ 
\phi _{f}(u)\cup \gamma _{f}(u),\;u\in U\cap V%
\end{array}%
\right.
\end{equation*}%
\begin{equation*}
(\Theta \cup \Gamma )_{f}=\underset{u\in U\cup V}{\bigcup }(\phi \cup \gamma
)_{f}(u)
\end{equation*}%
where the notations are obvious and similar with those from the previous
theorem.
\end{theorem}

\section{Subsystem}

\begin{definition}
Let $f:U\rightarrow P^{\ast }(S^{(n)})$ and $g:V\rightarrow P^{\ast
}(S^{(n)}),U,V\in P^{\ast }(S^{(m)})$ be two systems$.$ $f$ is called a
subsystem of $g$ if%
\begin{equation*}
U\subset V\text{ and }\forall u\in U,f(u)\subset g(u)
\end{equation*}
\end{definition}

\begin{remark}
The subsystem of a system represents a more precise model of the same
circuit, obtained perhaps after restricting the inputs set.

A special case in the inclusion $f\subset g$ is the one when $f$ is
uni-valued (it is called deterministic in this situation). This is
considered to be non-realistic in modeling.
\end{remark}

\begin{example}
Let $f$ be a system and we take some arbitrary $\mu \in \Theta _{0}.$ The
subsystem $f_{\mu }:U_{\mu }\rightarrow P^{\ast }(S^{(n)})$ defined by%
\begin{equation*}
U_{\mu }=\{u|u\in U,\mu \in \phi _{0}(u)\}
\end{equation*}%
\begin{equation*}
\forall u\in U_{\mu },f_{\mu }(u)=\{x|x\in f(u),x(-\infty +0)=\mu \}
\end{equation*}%
is called the restriction of $f$ at $\mu .$ The next property is satisfied:
for $\Theta _{0}=\{\mu ^{1},...,\mu ^{k}\},$ we have $f=f_{\mu ^{1}}\cup
...\cup f_{\mu ^{k}}$ (the union is not disjoint).
\end{example}

\begin{theorem}
Let $f:U\rightarrow P^{\ast }(S^{(n)}),$ $f_{1}:U_{1}\rightarrow P^{\ast
}(S^{(n)}),$ $g:V\rightarrow P^{\ast }(S^{(n)}),$ $g_{1}:V_{1}\rightarrow
P^{\ast }(S^{(n)})$ be some systems with $U,U_{1},V,V_{1}\in P^{\ast
}(S^{(m)}).$ If $f\subset f_{1},g\subset g_{1}$ and if $f\cap g$ exists,
then $f_{1}\cap g_{1}$ exists and the inclusion $f\cap g\subset f_{1}\cap
g_{1}$ is true.
\end{theorem}

\begin{proof}
We denote by $W$ the set from (\ref{ir_1}) and with $W_{1}$ the set%
\begin{equation*}
W_{1}=\{u|u\in U_{1}\cap V_{1},f_{1}(u)\cap g_{1}(u)\neq \emptyset \}
\end{equation*}%
From the fact that $U\subset U_{1}$, $\forall u\in U,f(u)\subset f_{1}(u),$ $%
V\subset V_{1}$, $\forall v\in V,g(v)\subset g_{1}(v)$ and $W\neq \emptyset $
we infer $W\subset W_{1},$ $W_{1}\neq \emptyset $ and furthermore we have $%
\forall u\in W,(f\cap g)(u)=f(u)\cap g(u)\subset f_{1}(u)\cap
g_{1}(u)=(f_{1}\cap g_{1})(u).$
\end{proof}

\begin{theorem}
We consider the systems $f:U\rightarrow P^{\ast }(S^{(n)}),$ $%
f_{1}:U_{1}\rightarrow P^{\ast }(S^{(n)}),$ $g:V\rightarrow P^{\ast
}(S^{(n)}),$ $g_{1}:V_{1}\rightarrow P^{\ast }(S^{(n)})$ with $%
U,U_{1},V,V_{1}\in P^{\ast }(S^{(m)}).$ If $f\subset f_{1},$ $g\subset g_{1}$
then $f\cup g\subset f_{1}\cup g_{1}.$
\end{theorem}

\begin{proof}
From $U\subset U_{1},V\subset V_{1}$ we infer that $U\cup V\subset U_{1}\cup
V_{1}.$ It is shown that $\forall u\in U\cup V,(f\cup g)(u)\subset
(f_{1}\cup g_{1})(u)$ is true in all the three situations $u\in U\setminus
V,u\in V\setminus U$ and $u\in U\cap V.$ For example if $u\in U\setminus V,$
then two possibilities exist:

- $u\in U_{1}\setminus V_{1},$ thus%
\begin{equation*}
(f\cup g)(u)=f(u)\subset f_{1}(u)=(f_{1}\cup g_{1})(u)
\end{equation*}

- $u\in U_{1}\cap V_{1},$ when%
\begin{equation*}
(f\cup g)(u)=f(u)\subset f_{1}(u)\subset f_{1}(u)\cup g_{1}(u)=(f_{1}\cup
g_{1})(u)
\end{equation*}%
is true. We observe that $u\in V_{1}\setminus U_{1}$ is impossible, since $%
u\notin U_{1}$ implies $u\notin U,$ contradiction.
\end{proof}

\section{Dual system}

\begin{notation}
For $u\in S^{(m)},$ we denote by $\overline{u}\in S^{(m)}$ the complement of 
$u$ satisfying%
\begin{equation*}
\forall t\in \mathbf{R},\overline{u}(t)=(\overline{u_{1}(t)},...,\overline{%
u_{m}(t)})
\end{equation*}
\end{notation}

\begin{definition}
\label{Def_23}The dual system of $f$ is the system $f^{\ast }:U^{\ast
}\rightarrow P^{\ast }(S^{(n)})$ defined in the next way%
\begin{equation*}
U^{\ast }=\{\overline{u}|u\in U\}
\end{equation*}%
\begin{equation*}
\forall u\in U^{\ast },f^{\ast }(u)=\{\overline{x}|x\in f(\overline{u})\}
\end{equation*}
\end{definition}

\begin{remark}
For any $u\in U^{\ast },$ $\overline{u}\in U$ and Definition \ref{Def_23} is
correct.

If $f$ models a circuit, then $f^{\ast }$ models the circuit that is
obtained from the previous one after the replacement of the OR gates with
AND gates and viceversa and respectively of the input and state tensions
with their complements (the complement of the 'HIGH' tension is by
definition the 'LOW' tension and viceversa).
\end{remark}

\begin{theorem}
\label{The87}If $f\cap g$ exists, then $(f\cap g)^{\ast },$ $f^{\ast }\cap
g^{\ast }$ exist and%
\begin{equation*}
(f\cap g)^{\ast }=f^{\ast }\cap g^{\ast }
\end{equation*}
\end{theorem}

\begin{proof}
We denote by $W$ the domain (\ref{ir_1}) of $f\cap g$. The domain of $(f\cap
g)^{\ast }$ is $W^{\ast }$ and the domain $W_{1}$ of $f^{\ast }\cap g^{\ast
} $ is:%
\begin{equation*}
W_{1}=\{u|u\in U^{\ast }\cap V^{\ast },f^{\ast }(u)\cap g^{\ast }(u)\neq
\emptyset \}=
\end{equation*}%
\begin{equation*}
=\{u|\overline{u}\in U\cap V,\{\overline{x}|x\in f(\overline{u})\}\cap \{%
\overline{x}|x\in g(\overline{u})\}\neq \emptyset \}=
\end{equation*}%
\begin{equation*}
=\{\overline{u}|u\in U\cap V,\{\overline{x}|x\in f(u)\}\cap \{\overline{x}%
|x\in g(u)\}\neq \emptyset \}=
\end{equation*}%
\begin{equation*}
=\{\overline{u}|u\in U\cap V,\{x|x\in f(u)\}\cap \{x|x\in g(u)\}\neq
\emptyset \}=W^{\ast }
\end{equation*}%
Moreover, for any $u\in W^{\ast }$ we infer%
\begin{equation*}
(f\cap g)^{\ast }(u)=\{\overline{x}|x\in (f\cap g)(\overline{u})\}=\{%
\overline{x}|x\in f(\overline{u})\cap g(\overline{u})\}=
\end{equation*}%
\begin{equation*}
=\{\overline{x}|x\in f(\overline{u})\}\cap \{\overline{x}|x\in g(\overline{u}%
)\}=f^{\ast }(u)\cap g^{\ast }(u)=(f^{\ast }\cap g^{\ast })(u)
\end{equation*}
\end{proof}

\begin{theorem}
\label{The97}We have%
\begin{equation*}
(f\cup g)^{\ast }=f^{\ast }\cup g^{\ast }
\end{equation*}
\end{theorem}

\begin{proof}
We remark that the equal domains of the two systems are $(U\cup V)^{\ast
}=U^{\ast }\cup V^{\ast }.$ Let $u\in U^{\ast }\cup V^{\ast }$ be an
arbitrary input$.$ If $u\in U^{\ast }\setminus V^{\ast },$ then $f^{\ast
}(u)=(f^{\ast }\cup g^{\ast })(u)$ and the fact that $\overline{u}\in
U\setminus V$ implies $(f\cup g)(\overline{u})=f(\overline{u}),$ thus%
\begin{equation*}
(f\cup g)^{\ast }(u)=\{\overline{x}|x\in (f\cup g)(\overline{u})\}=\{%
\overline{x}|x\in f(\overline{u})\}=f^{\ast }(u)=(f^{\ast }\cup g^{\ast })(u)
\end{equation*}%
If $u\in V^{\ast }\setminus U^{\ast }$, the situation is similar. We presume
in this moment that $u\in U^{\ast }\cap V^{\ast },$ implying $f^{\ast
}(u)\cup g^{\ast }(u)=(f^{\ast }\cup g^{\ast })(u),$ $\overline{u}\in U\cap
V,(f\cup g)(\overline{u})=f(\overline{u})\cup g(\overline{u})$ and we have:%
\begin{equation*}
(f\cup g)^{\ast }(u)=\{\overline{x}|x\in (f\cup g)(\overline{u})\}=\{%
\overline{x}|x\in f(\overline{u})\cup g(\overline{u})\}=
\end{equation*}%
\begin{equation*}
=\{\overline{x}|x\in f(\overline{u})\}\cup \{\overline{x}|x\in g(\overline{u}%
)\}=f^{\ast }(u)\cup g^{\ast }(u)=(f^{\ast }\cup g^{\ast })(u)
\end{equation*}%
In all the three cases the statement of the theorem was proved to be true.
\end{proof}

\section{Inverse system}

\begin{definition}
The inverse system of $f$ is defined by $f^{-1}:X\rightarrow P^{\ast
}(S^{m)}),$%
\begin{equation*}
X=\underset{u\in U}{\bigcup }f(u)
\end{equation*}%
\begin{equation*}
\forall x\in X,f^{-1}(x)=\{u|u\in U,x\in f(u)\}
\end{equation*}
\end{definition}

\begin{remark}
The inputs and the states of $f$ become states and inputs of $f^{-1},$
meaning that $f^{-1}$ inverts the causes and the effects in modeling: its
aim is to answer the question 'given an effect $x$, which are the causes $u$
producing it?'
\end{remark}

\begin{theorem}
Let $f:U\rightarrow P^{\ast }(S^{(n)}),g:V\rightarrow P^{\ast
}(S^{(n)}),U,V\in P^{\ast }(S^{(m)})$ be some systems$.$ If $\exists u\in
U\cap V,f(u)\cap g(u)\neq \emptyset ,$ then the systems $(f\cap g)^{-1},$ $%
f^{-1}\cap g^{-1}$ exist and they have the same domain:%
\begin{equation*}
Y=\underset{u\in W}{\bigcup }(f(u)\cap g(u))
\end{equation*}%
Furthermore, we have%
\begin{equation*}
(f\cap g)^{-1}=f^{-1}\cap g^{-1}
\end{equation*}
\end{theorem}

\begin{proof}
$Y$ is obviously the domain of $(f\cap g)^{-1}.$ We can write%
\begin{equation*}
Y=\underset{u\in U\cap V}{\bigcup }(f(u)\cap g(u))=\{x|\exists u\in U\cap
V,x\in f(u)\cap g(u)\}=
\end{equation*}%
\begin{equation*}
=\{x|x\in \underset{v\in U}{\bigcup }f(v)\cap \underset{v\in V}{\bigcup }%
g(v),\exists u,u\in U,x\in f(u)\text{ and }u\in V,x\in g(u)\}=
\end{equation*}%
\begin{equation*}
=\{x|x\in \underset{v\in U}{\bigcup }f(v)\cap \underset{v\in V}{\bigcup }%
g(v),\exists u,u\in f^{-1}(x)\text{ and }u\in g^{-1}(x)\}=
\end{equation*}%
\begin{equation*}
=\{x|x\in \underset{v\in U}{\bigcup }f(v)\cap \underset{v\in V}{\bigcup }%
g(v),f^{-1}(x)\cap g^{-1}(x)\neq \emptyset \}
\end{equation*}%
thus $Y$ is the domain of $f^{-1}\cap g^{-1}$ too$.$ We have $\forall x\in
Y, $%
\begin{equation*}
(f\cap g)^{-1}(x)=\{u|u\in U\cap V,x\in (f\cap g)(u)\}=\{u|u\in U\cap V,x\in
f(u)\cap g(u)\}=
\end{equation*}%
\begin{equation*}
=\{u|u\in U\cap V,x\in f(u)\}\cap \{u|u\in U\cap V,x\in g(u)\}=
\end{equation*}%
\begin{equation*}
=(\{u|u\in U\setminus V,x\in f(u)\}\cup \{u|u\in U\cap V,x\in f(u)\})\cap
\end{equation*}%
\begin{equation*}
\cap (\{u|u\in V\setminus U,x\in g(u)\}\cup \{u|u\in U\cap V,x\in g(u)\})=
\end{equation*}%
\begin{equation*}
=\{u|u\in U,x\in f(u)\}\cap \{u|u\in V,x\in g(u)\}=f^{-1}(x)\cap
g^{-1}(x)=(f^{-1}\cap g^{-1})(x)
\end{equation*}
\end{proof}

\begin{theorem}
\label{The18}We consider the systems $f:U\rightarrow P^{\ast
}(S^{(n)}),g:V\rightarrow P^{\ast }(S^{(n)}),U,V\in P^{\ast }(S^{(m)}).$ The
systems $(f\cup g)^{-1},$ $f^{-1}\cup g^{-1}$ have the domain equal with%
\begin{equation*}
Y^{\prime }=\underset{u\in U}{\bigcup }f(u)\cup \underset{u\in V}{\bigcup }%
g(u)
\end{equation*}%
and the next equality is true%
\begin{equation*}
(f\cup g)^{-1}=f^{-1}\cup g^{-1}
\end{equation*}
\end{theorem}

\begin{proof}
The domain of $(f\cup g)^{-1}$ is 
\begin{equation*}
\underset{u\in U\cup V}{\bigcup }(f\cup g)(u)=\underset{u\in U\setminus V}{%
\bigcup }(f\cup g)(u)\cup \underset{u\in U\cap V}{\bigcup }(f\cup g)(u)\cup 
\underset{u\in V\setminus U}{\bigcup }(f\cup g)(u)=
\end{equation*}%
\begin{equation*}
=\underset{u\in U\setminus V}{\bigcup }f(u)\cup \underset{u\in U\cap V}{%
\bigcup }(f(u)\cup g(u))\cup \underset{u\in V\setminus U}{\bigcup }g(u)=
\end{equation*}%
\begin{equation*}
=\underset{u\in U\setminus V}{\bigcup }f(u)\cup \underset{u\in U\cap V}{%
\bigcup }f(u)\cup \underset{u\in U\cap V}{\bigcup }g(u)\cup \underset{u\in
V\setminus U}{\bigcup }g(u)=\underset{u\in U}{\bigcup }f(u)\cup \underset{%
u\in V}{\bigcup }g(u)
\end{equation*}%
and it coincides with $Y^{\prime }$, that is obviously the domain of $%
f^{-1}\cup g^{-1}.$ For any $x\in Y^{\prime }$ we have:%
\begin{equation*}
(f\cup g)^{-1}(x)=\{u|u\in U\cup V,x\in (f\cup g)(u)\}=\{u|u\in U\setminus
V,x\in f(u)\}\cup
\end{equation*}%
\begin{equation*}
\cup \{u|u\in V\setminus U,x\in g(u)\}\cup \{u|u\in U\cap V,x\in f(u)\}\cup
\{u|u\in U\cap V,x\in g(u)\}=
\end{equation*}%
\begin{equation*}
=\{u|u\in U,x\in f(u)\}\cup \{u|u\in V,x\in g(u)\}=
\end{equation*}%
\begin{equation*}
=\left\{ 
\begin{array}{c}
f^{-1}(x),x\in \underset{u\in U}{\bigcup }f(u)\setminus \underset{u\in V}{%
\bigcup }g(u) \\ 
g^{-1}(x),x\in \underset{u\in V}{\bigcup }g(u)\setminus \underset{u\in U}{%
\bigcup }f(u) \\ 
f^{-1}(x)\cup g^{-1}(x),x\in \underset{u\in U}{\bigcup }f(u)\cap \underset{%
u\in V}{\bigcup }g(u)%
\end{array}%
\right. =(f^{-1}\cup g^{-1})(x)
\end{equation*}
\end{proof}

\section{Cartesian product}

\begin{definition}
Let $u\in S^{(m)},u^{\prime }\in S^{(m^{\prime })}$ be two signals$.$ We
define the Cartesian product $u\times u^{\prime }\in S^{(m+m^{\prime })}$ of
the functions $u$ and $u^{\prime }$ by%
\begin{equation*}
\forall t\in \mathbf{R},(u\times u^{\prime
})(t)=(u_{1}(t),...,u_{m}(t),u_{1}^{\prime }(t),...,u_{m^{\prime }}^{\prime
}(t))
\end{equation*}
\end{definition}

\begin{definition}
For any sets $U\in P^{\ast }(S^{(m)}),U^{\prime }\in P^{\ast }(S^{(m^{\prime
})})$ we define the Cartesian product $U\times U^{\prime }\in P^{\ast
}(S^{(m+m^{\prime })}),$%
\begin{equation*}
U\times U^{\prime }=\{u\times u^{\prime }|u\in U,u^{\prime }\in U^{\prime }\}
\end{equation*}
\end{definition}

\begin{definition}
The Cartesian product of the systems $f$ and $f^{\prime }:U^{\prime
}\rightarrow P^{\ast }(S^{(n^{\prime })}),U^{\prime }\in P^{\ast
}(S^{(m^{\prime })})$ is $f\times f^{\prime }:U\times U^{\prime }\rightarrow
P^{\ast }(S^{(n+n^{\prime })}),$%
\begin{equation*}
\forall u\times u^{\prime }\in U\times U^{\prime },(f\times f^{\prime
})(u\times u^{\prime })=f(u)\times f^{\prime }(u^{\prime })
\end{equation*}
\end{definition}

\begin{remark}
The Cartesian product of the systems models two circuits that are not
interconnected and run under different inputs.
\end{remark}

\begin{theorem}
\label{The81}Let $f:U\rightarrow P^{\ast }(S^{(n)}),$ $g:V\rightarrow
P^{\ast }(S^{(n)}),$ $U,V\in P^{\ast }(S^{(m)})$ and $f^{\prime }:U^{\prime
}\rightarrow P^{\ast }(S^{(n^{\prime })}),$ $U^{\prime }\in P^{\ast
}(S^{(m^{\prime })})$ be three systems$.$ If $\exists u\in U\cap V,$ $%
f(u)\cap g(u)\neq \emptyset $ then the systems $(f\cap g)\times f^{\prime },$
$(f\times f^{\prime })\cap (g\times f^{\prime })$ are defined and $W\times
U^{\prime }$ is their common domain, where we have used again the notation%
\begin{equation*}
W=\{u|u\in U\cap V,f(u)\cap g(u)\neq \emptyset \}
\end{equation*}%
The next equality is true%
\begin{equation*}
(f\cap g)\times f^{\prime }=(f\times f^{\prime })\cap (g\times f^{\prime })
\end{equation*}
\end{theorem}

\begin{proof}
We show that $W\times U^{\prime },$ that is the domain of $(f\cap g)\times
f^{\prime },$ is also the domain of $(f\times f^{\prime })\cap (g\times
f^{\prime }):$%
\begin{equation*}
W\times U^{\prime }=\{u\times u^{\prime }|u\in W,u^{\prime }\in U^{\prime
}\}=
\end{equation*}%
\begin{equation*}
=\{u\times u^{\prime }|u\in U\cap V,u^{\prime }\in U^{\prime },f(u)\cap
g(u)\neq \emptyset \text{ and }f^{\prime }(u^{\prime })\neq \emptyset \}=
\end{equation*}%
\begin{equation*}
=\{u\times u^{\prime }|u\times u^{\prime }\in (U\cap V)\times U^{\prime
},(f(u)\times f^{\prime }(u^{\prime }))\cap (g(u)\times f^{\prime
}(u^{\prime }))\neq \emptyset \}=
\end{equation*}%
\begin{equation*}
=\{u\times u^{\prime }|u\times u^{\prime }\in (U\times U^{\prime })\cap
(V\times U^{\prime }),(f\times f^{\prime })(u\times u^{\prime })\cap
(g\times f^{\prime })(u\times u^{\prime })\neq \emptyset \}
\end{equation*}%
Furthermore for any $u\times u^{\prime }\in W\times U^{\prime }$ we have%
\begin{equation*}
((f\cap g)\times f^{\prime })(u\times u^{\prime })=(f\cap g)(u)\times
f^{\prime }(u^{\prime })=(f(u)\cap g(u))\times f^{\prime }(u^{\prime })=
\end{equation*}%
\begin{equation*}
=(f(u)\times f^{\prime }(u^{\prime }))\cap (g(u)\times f^{\prime }(u^{\prime
}))=(f\times f^{\prime })(u\times u^{\prime })\cap (g\times f^{\prime
})(u\times u^{\prime })=
\end{equation*}%
\begin{equation*}
=((f\times f^{\prime })\cap (g\times f^{\prime }))(u\times u^{\prime })
\end{equation*}
\end{proof}

\begin{theorem}
Let $f:U\rightarrow P^{\ast }(S^{(n)}),$ $g:V\rightarrow P^{\ast }(S^{(n)}),$
$U,V\in P^{\ast }(S^{(m)})$ and $f^{\prime }:U^{\prime }\rightarrow P^{\ast
}(S^{(n^{\prime })}),$ $U^{\prime }\in P^{\ast }(S^{(m^{\prime })})$ be some
systems$.$ The common domain of $(f\cup g)\times f^{\prime },$ $(f\times
f^{\prime })\cup (g\times f^{\prime })$ is $(U\cup V)\times U^{\prime
}=(U\times U^{\prime })\cup (V\times U^{\prime })$ and the next equality
holds%
\begin{equation*}
(f\cup g)\times f^{\prime }=(f\times f^{\prime })\cup (g\times f^{\prime })
\end{equation*}
\end{theorem}

\begin{proof}
$\forall u\times u^{\prime }\in (U\cup V)\times U^{\prime }$ we have one of
the next possibilities:

\emph{Case} $u\times u^{\prime }\in (U\setminus V)\times U^{\prime
}=(U\times U^{\prime })\setminus (V\times U^{\prime })$%
\begin{equation*}
((f\cup g)\times f^{\prime })(u\times u^{\prime })=(f\cup g)(u)\times
f^{\prime }(u^{\prime })=f(u)\times f^{\prime }(u^{\prime })=(f\times
f^{\prime })(u\times u^{\prime })=
\end{equation*}%
\begin{equation*}
=((f\times f^{\prime })\cup (g\times f^{\prime }))(u\times u^{\prime })
\end{equation*}

\emph{Case} $u\times u^{\prime }\in (V\setminus U)\times U^{\prime }$ is
similar

\emph{Case} $u\times u^{\prime }\in (U\cap V)\times U^{\prime }=(U\times
U^{\prime })\cap (V\times U^{\prime })$%
\begin{equation*}
((f\cup g)\times f^{\prime })(u\times u^{\prime })=(f\cup g)(u)\times
f^{\prime }(u^{\prime })=(f(u)\cup g(u))\times f^{\prime }(u^{\prime })=
\end{equation*}%
\begin{equation*}
=(f(u)\times f^{\prime }(u^{\prime }))\cup (g(u)\times f^{\prime }(u^{\prime
}))=(f\times f^{\prime })(u\times u^{\prime })\cup (g\times f^{\prime
})(u\times u^{\prime })=
\end{equation*}%
\begin{equation*}
=((f\times f^{\prime })\cup (g\times f^{\prime }))(u\times u^{\prime })
\end{equation*}
\end{proof}

\section{Parallel connection}

\begin{definition}
The parallel connection of $f$ with $f_{1}^{\prime }:U_{1}^{\prime
}\rightarrow P^{\ast }(S^{(n^{\prime })}),U_{1}^{\prime }\in P^{\ast
}(S^{(m)})$ is $(f,f_{1}^{\prime }):U\cap U_{1}^{\prime }\rightarrow P^{\ast
}(S^{(n+n^{\prime })}),$%
\begin{equation*}
\forall u\in U\cap U_{1}^{\prime },(f,f_{1}^{\prime })(u)=(f\times
f_{1}^{\prime })(u\times u)
\end{equation*}
\end{definition}

\begin{remark}
The parallel connection models two circuits that are not interconnected and
run under the same input.
\end{remark}

\begin{theorem}
\label{The82}We consider the systems $f:U\rightarrow P^{\ast }(S^{(n)}),$ $%
g:V\rightarrow P^{\ast }(S^{(n)}),$ $f_{1}^{\prime }:U_{1}^{\prime
}\rightarrow P^{\ast }(S^{(n^{\prime })}),$ with $U,V,U_{1}^{\prime }\in
P^{\ast }(S^{(m)})$. We presume that $\exists u\in U\cap V\cap U_{1}^{\prime
}$ so that $f(u)\cap g(u)\neq \emptyset .$ Then the set%
\begin{equation*}
W^{\prime }=\{u|u\in U\cap V\cap U_{1}^{\prime },f(u)\cap g(u)\neq \emptyset
\}
\end{equation*}%
is the domain of the systems $(f\cap g,f_{1}^{\prime }),$ $(f,f_{1}^{\prime
})\cap (g,f_{1}^{\prime })$ and the next equality holds%
\begin{equation*}
(f\cap g,f_{1}^{\prime })=(f,f_{1}^{\prime })\cap (g,f_{1}^{\prime })
\end{equation*}
\end{theorem}

\begin{proof}
We observe that $W^{\prime }$ is non-empty, it is the domain of $(f\cap
g,f_{1}^{\prime })$ and we show that it is also the domain of $%
(f,f_{1}^{\prime })\cap (g,f_{1}^{\prime }).$ We denote by%
\begin{equation*}
W"=\{u|u\in (U\cap U_{1}^{\prime })\cap (V\cap U_{1}^{\prime
}),(f,f_{1}^{\prime })(u)\cap (g,f_{1}^{\prime })(u)\neq \emptyset \}
\end{equation*}%
the domain of $(f,f_{1}^{\prime })\cap (g,f_{1}^{\prime })$ for which we have%
\begin{equation*}
W"=\{u|u\in U\cap V\cap U_{1}^{\prime },(f(u)\times f_{1}^{\prime }(u))\cap
(g(u)\times f_{1}^{\prime }(u))\neq \emptyset \}=
\end{equation*}%
\begin{equation*}
=\{u|u\in U\cap V\cap U_{1}^{\prime },(f(u)\cap g(u))\times f_{1}^{\prime
}(u)\neq \emptyset \}=
\end{equation*}%
\begin{equation*}
=\{u|u\in U\cap V\cap U_{1}^{\prime },f(u)\cap g(u)\neq \emptyset \}
\end{equation*}%
thus $W"=W^{\prime }.$ For any $u\in W^{\prime }$ we have:%
\begin{equation*}
(f\cap g,f_{1}^{\prime })(u)=((f\cap g)\times f_{1}^{\prime })(u\times u)%
\overset{Theorem\text{ }\ref{The81}}{=}((f\times f_{1}^{\prime })\cap
(g\times f_{1}^{\prime }))(u\times u)=
\end{equation*}%
\begin{equation*}
=(f\times f_{1}^{\prime })(u\times u)\cap (g\times f_{1}^{\prime })(u\times
u)=(f,f_{1}^{\prime })(u)\cap (g,f_{1}^{\prime })(u)=((f,f_{1}^{\prime
})\cap (g,f_{1}^{\prime }))(u)
\end{equation*}
\end{proof}

\begin{remark}
A similar result with the one from Theorem \ref{The81} states the truth of
the formula%
\begin{equation*}
f\times (f^{\prime }\cap g^{\prime })=(f\times f^{\prime })\cap (f\times
g^{\prime })
\end{equation*}%
and then from Theorem \ref{The81} we get the next property%
\begin{equation*}
(f\cap g)\times (f^{\prime }\cap g^{\prime })=(f\times f^{\prime })\cap
(f\times g^{\prime })\cap (g\times f^{\prime })\cap (g\times g^{\prime })
\end{equation*}%
Like in Theorem \ref{The82} we can prove that%
\begin{equation*}
(f,f_{1}^{\prime }\cap g_{1}^{\prime })=(f,f_{1}^{\prime })\cap
(f,g_{1}^{\prime })
\end{equation*}%
is true and then from Theorem \ref{The82} we obtain%
\begin{equation*}
(f\cap g,f_{1}^{\prime }\cap g_{1}^{\prime })=(f,f_{1}^{\prime })\cap
(f,g_{1}^{\prime })\cap (g,f_{1}^{\prime })\cap (g,g_{1}^{\prime })
\end{equation*}
\end{remark}

\begin{theorem}
Let $f:U\rightarrow P^{\ast }(S^{(n)}),$ $g:V\rightarrow P^{\ast }(S^{(n)}),$
$f_{1}^{\prime }:U_{1}^{\prime }\rightarrow P^{\ast }(S^{(n^{\prime })})$ be
three systems with $U,V,U_{1}^{\prime }\in P^{\ast }(S^{(m)})$. If $U\cap
U_{1}^{\prime }\neq \emptyset ,$ $V\cap U_{1}^{\prime }\neq \emptyset $,
then the common domain of the systems $(f\cup g,f_{1}^{\prime }),$ $%
(f,f_{1}^{\prime })\cup (g,f_{1}^{\prime })$ is $(U\cup V)\cap U_{1}^{\prime
}=(U\cap U_{1}^{\prime })\cup (V\cap U_{1}^{\prime })$ and we have%
\begin{equation*}
(f\cup g,f_{1}^{\prime })=(f,f_{1}^{\prime })\cup (g,f_{1}^{\prime })
\end{equation*}
\end{theorem}

\begin{remark}
We observe the truth of the formulas%
\begin{equation*}
f\times (f^{\prime }\cup g^{\prime })=(f\times f^{\prime })\cup (f\times
g^{\prime })
\end{equation*}%
\begin{equation*}
(f\cup g)\times (f^{\prime }\cup g^{\prime })=(f\times f^{\prime })\cup
(f\times g^{\prime })\cup (g\times f^{\prime })\cup (g\times g^{\prime })
\end{equation*}%
and respectively of the formulas%
\begin{equation*}
(f,f_{1}^{\prime }\cup g_{1}^{\prime })=(f,f_{1}^{\prime })\cup
(f,g_{1}^{\prime })
\end{equation*}%
\begin{equation*}
(f\cup g,f_{1}^{\prime }\cup g_{1}^{\prime })=(f,f_{1}^{\prime })\cup
(f,g_{1}^{\prime })\cup (g,f_{1}^{\prime })\cup (g,g_{1}^{\prime })
\end{equation*}
\end{remark}

\section{Serial connection}

\begin{definition}
The serial connection of $h:X\rightarrow P^{\ast }(S^{(p)}),X\in P^{\ast
}(S^{(n)})$ with $f:U\rightarrow P^{\ast }(S^{(n)}),U\in P^{\ast }(S^{(m)})$
is defined whenever $\underset{u\in U}{\bigcup }f(u)\subset X$ by\footnote{%
we show a more general definition of the serial connection that was used in
previous works: the request $\underset{u\in U}{\bigcup }f(u)\subset X$ is
replaced by $\exists u\in U,f(u)\cap X\neq \emptyset $ and $h\circ
f:Z\rightarrow P^{\ast }(S^{(p)})$ is defined by%
\begin{equation*}
Z=\{u|u\in U,f(u)\cap X\neq \emptyset \}
\end{equation*}%
\begin{equation*}
\forall u\in Z,(h\circ f)(u)=\underset{x\in f(u)\cap X}{\bigcup }h(x)
\end{equation*}%
} $h\circ f:U\rightarrow P^{\ast }(S^{(p)}),$%
\begin{equation*}
\forall u\in U,(h\circ f)(u)=\underset{x\in f(u)}{\bigcup }h(x)
\end{equation*}
\end{definition}

\begin{remark}
The serial connection of the systems models two circuits connected in
cascade and it coincides with the usual composition of the multi-valued
functions.
\end{remark}

\begin{theorem}
We consider the systems $f:U\rightarrow P^{\ast }(S^{(n)}),$ $g:V\rightarrow
P^{\ast }(S^{(n)}),$ $U,V\in P^{\ast }(S^{(m)})$ and $h:X\rightarrow P^{\ast
}(S^{(p)}),$ $h_{1}:X_{1}\rightarrow P^{\ast }(S^{(p)}),$ $X,X_{1}\in
P^{\ast }(S^{(n)}).$

a) If $\underset{u\in U}{\bigcup }f(u)\subset X,\underset{u\in V}{\bigcup }%
g(u)\subset X$ and $\exists u\in U\cap V,f(u)\cap g(u)\neq \emptyset $ then
the sets%
\begin{equation*}
W=\{u|u\in U\cap V,f(u)\cap g(u)\neq \emptyset \}
\end{equation*}%
\begin{equation*}
W_{1}=\{u|u\in U\cap V,\underset{x\in f(u)}{\bigcup }h(x)\cap \underset{x\in
g(u)}{\bigcup }h(x)\neq \emptyset \}
\end{equation*}%
are non-empty and represent the domains of the systems $h\circ (f\cap g),$ $%
(h\circ f)\cap (h\circ g).$ We have%
\begin{equation*}
h\circ (f\cap g)\subset (h\circ f)\cap (h\circ g)
\end{equation*}

b) We ask that $\underset{u\in U}{\bigcup }f(u)\subset \{x|x\in X\cap
X_{1},h(x)\cap h_{1}(x)\neq \emptyset \}.$ $U$ is the domain of the systems $%
(h\cap h_{1})\circ f,$ $(h\circ f)\cap (h_{1}\circ f)$ and the next
inclusion is true:%
\begin{equation*}
(h\cap h_{1})\circ f\subset (h\circ f)\cap (h_{1}\circ f)
\end{equation*}
\end{theorem}

\begin{proof}
a) From the hypothesis $f\cap g$ is defined and has the domain $W.$ As%
\begin{equation*}
\underset{u\in W}{\bigcup }(f\cap g)(u)\subset \underset{u\in W}{\bigcup }%
f(u)\subset \underset{u\in U}{\bigcup }f(u)\subset X
\end{equation*}%
we have obtained that $h\circ (f\cap g)$ is defined and has the domain $W.$

From the same hypothesis $h\circ f$ and $h\circ g$ are defined and have the
domains $U,V.$ Because $\emptyset \neq W\subset W_{1},$ the system $(h\circ
f)\cap (h\circ g)$ is defined and has the domain $W_{1}.$

$\forall u\in W$ we get%
\begin{equation*}
\underset{x\in f(u)\cap g(u)}{\bigcup }h(x)\subset \underset{x\in f(u)}{%
\bigcup }h(x),\text{ }\underset{x\in f(u)\cap g(u)}{\bigcup }h(x)\subset 
\underset{x\in g(u)}{\bigcup }h(x)
\end{equation*}%
from where%
\begin{equation*}
\underset{x\in f(u)\cap g(u)}{\bigcup }h(x)\subset \underset{x\in f(u)}{%
\bigcup }h(x)\cap \underset{x\in g(u)}{\bigcup }h(x)
\end{equation*}%
and we conclude that $\forall u\in W,$%
\begin{equation*}
(h\circ (f\cap g))(u)=\underset{x\in f(u)\cap g(u)}{\bigcup }h(x)\subset 
\underset{x\in f(u)}{\bigcup }h(x)\cap \underset{x\in g(u)}{\bigcup }h(x)=
\end{equation*}%
\begin{equation*}
=(h\circ f)(u)\cap (h\circ g)(u)=((h\circ f)\cap (h\circ g))(u)
\end{equation*}

b) The hypothesis $\underset{u\in U}{\bigcup }f(u)\subset \{x|x\in X\cap
X_{1},h(x)\cap h_{1}(x)\neq \emptyset \}$ states that the domain $\{x|x\in
X\cap X_{1},h(x)\cap h_{1}(x)\neq \emptyset \}$ of $h\cap h_{1}$ is
non-empty and that $(h\cap h_{1})\circ f$ is defined. From the hypothesis we
infer that $\underset{u\in U}{\bigcup }f(u)\subset X,\underset{u\in U}{%
\bigcup }f(u)\subset X_{1}$ and $h\circ f,h_{1}\circ f$ are defined
themselves. The domain of $(h\cap h_{1})\circ f$ is $U.$ Moreover from $%
\forall u\in U,\forall x\in f(u),h(x)\cap h_{1}(x)\neq \emptyset $ we
conclude that the domain $\{u|u\in U,\underset{x\in f(u)}{\bigcup }h(x)\cap 
\underset{x\in f(u)}{\bigcup }h_{1}(x)\neq \emptyset \}$ of $(h\circ f)\cap
(h_{1}\circ f)$ is equal with $U$ too$.$

Let $u\in U$ be arbitrary and fixed$.$ From%
\begin{equation*}
\underset{x\in f(u)}{\bigcup }(h\cap h_{1})(x)\subset \underset{x\in f(u)}{%
\bigcup }h(x),\underset{x\in f(u)}{\bigcup }(h\cap h_{1})(x)\subset \underset%
{x\in f(u)}{\bigcup }h_{1}(x)
\end{equation*}%
we get%
\begin{equation*}
\underset{x\in f(u)}{\bigcup }(h\cap h_{1})(x)\subset \underset{x\in f(u)}{%
\bigcup }h(x)\cap \underset{x\in f(u)}{\bigcup }h_{1}(x)
\end{equation*}%
and eventually we obtain%
\begin{equation*}
((h\cap h_{1})\circ f)(u)=\underset{x\in f(u)}{\bigcup }(h\cap
h_{1})(x)\subset
\end{equation*}%
\begin{equation*}
\subset \underset{x\in f(u)}{\bigcup }h(x)\cap \underset{x\in f(u)}{\bigcup }%
h_{1}(x)=(h\circ f)(u)\cap (h_{1}\circ f)(u)=((h\circ f)\cap (h_{1}\circ
f))(u)
\end{equation*}
\end{proof}

\begin{theorem}
We have the systems $f:U\rightarrow P^{\ast }(S^{(n)}),$ $g:V\rightarrow
P^{\ast }(S^{(n)}),$ $U,V\in P^{\ast }(S^{(m)})$ and $h:X\rightarrow P^{\ast
}(S^{(p)}),$ $h_{1}:X_{1}\rightarrow P^{\ast }(S^{(p)}),$ $X,X_{1}\in
P^{\ast }(S^{(n)}).$

a) We presume that $\underset{u\in U}{\bigcup }f(u)\subset X,\underset{u\in V%
}{\bigcup }g(u)\subset X;$ the set $U\cup V$ is the common domain of $h\circ
(f\cup g),$ $(h\circ f)\cup (h\circ g)$ and the next equality is true%
\begin{equation*}
h\circ (f\cup g)=(h\circ f)\cup (h\circ g)
\end{equation*}

b) If $\underset{u\in U}{\bigcup }f(u)\subset X,\underset{u\in U}{\bigcup }%
f(u)\subset X_{1}$ then $(h\cup h_{1})\circ f,$ $(h\circ f)\cup (h_{1}\circ
f)$ have the domain $U$ and%
\begin{equation*}
(h\cup h_{1})\circ f=(h\circ f)\cup (h_{1}\circ f)
\end{equation*}
\end{theorem}

\begin{proof}
a) The systems $h\circ f$ and $h\circ g$ are defined from the hypothesis and
because (see the proof of Theorem \ref{The18})%
\begin{equation*}
\underset{u\in U\cup V}{\bigcup }(f\cup g)(u)=\underset{u\in U}{\bigcup }%
f(u)\cup \underset{u\in V}{\bigcup }g(u)\subset X
\end{equation*}%
we infer that $h\circ (f\cup g)$ is defined too. The common domain of $%
h\circ (f\cup g)$ and $(h\circ f)\cup (h\circ g)$ is $U\cup V.$

Let $u\in U\cup V$ be arbitrary. We can prove the statement of the theorem
in the three cases: $u\in (U\setminus V),$ $u\in (V\setminus U),$ $u\in
(U\cap V)$. For example in the last case we have:%
\begin{equation*}
(h\circ (f\cup g))(u)=\underset{x\in (f\cup g)(u)}{\bigcup }h(x)=\underset{%
x\in f(u)\cup g(u)}{\bigcup }h(x)=
\end{equation*}%
\begin{equation*}
=\underset{x\in f(u)}{\bigcup }h(x)\cup \underset{x\in g(u)}{\bigcup }%
h(x)=(h\circ f)(u)\cup (h\circ g)(u)=((h\circ f)\cup (h\circ g))(u)
\end{equation*}

b) The hypothesis implies $\underset{u\in U}{\bigcup }f(u)\subset X\cup
X_{1} $ thus $(h\cup h_{1})\circ f$ is defined and on the other hand $h\circ
f$ and $h_{1}\circ f$ are defined too. The systems $(h\cup h_{1})\circ f$, $%
(h\circ f)\cup (h_{1}\circ f)$ have the same domain $U=U\cup U.$

For any $u\in U$ fixed, we have%
\begin{equation*}
f(u)=f(u)\cap (X\cup X_{1})=f(u)\cap ((X\setminus X_{1})\cup (X_{1}\setminus
X)\cup (X\cap X_{1}))=
\end{equation*}%
\begin{equation*}
=(f(u)\cap (X\setminus X_{1}))\cup (f(u)\cap (X_{1}\setminus X))\cup
(f(u)\cap (X\cap X_{1}))
\end{equation*}%
thus%
\begin{equation*}
((h\cup h_{1})\circ f)(u)=\underset{x\in f(u)}{\bigcup }(h\cup h_{1})(x)=
\end{equation*}%
\begin{equation*}
=\underset{x\in (f(u)\cap (X\setminus X_{1}))\cup (f(u)\cap (X_{1}\setminus
X))\cup (f(u)\cap (X\cap X_{1}))}{\bigcup }(h\cup h_{1})(x)=
\end{equation*}%
\begin{equation*}
=\underset{x\in f(u)\cap (X\setminus X_{1})}{\bigcup }(h\cup h_{1})(x)\cup 
\underset{x\in f(u)\cap (X_{1}\setminus X)}{\bigcup }(h\cup h_{1})(x)\cup 
\underset{x\in f(u)\cap X\cap X_{1}}{\bigcup }(h\cup h_{1})(x)=
\end{equation*}%
\begin{equation*}
=\underset{x\in f(u)\cap (X\setminus X_{1})}{\bigcup }h(x)\cup \underset{%
x\in f(u)\cap (X_{1}\setminus X)}{\bigcup }h_{1}(x)\cup \underset{x\in
f(u)\cap X\cap X_{1}}{\bigcup }h(x)\cup \underset{x\in f(u)\cap X\cap X_{1}}{%
\bigcup }h_{1}(x)=
\end{equation*}%
\begin{equation*}
=\underset{x\in (f(u)\cap (X\setminus X_{1}))\cup (f(u)\cap X\cap X_{1})}{%
\bigcup }h(x)\cup \underset{x\in (f(u)\cap (X_{1}\setminus X))\cup (f(u)\cap
X\cap X_{1})}{\bigcup }h_{1}(x)=
\end{equation*}%
\begin{equation*}
=\underset{x\in f(u)\cap X}{\bigcup }h(x)\cup \underset{x\in f(u)\cap X_{1}}{%
\bigcup }h_{1}(x)=\underset{x\in f(u)}{\bigcup }h(x)\cup \underset{x\in f(u)}%
{\bigcup }h_{1}(x)=
\end{equation*}%
\begin{equation*}
=(h\circ f)(u)\cup (h_{1}\circ f)(u)=((h\circ f)\cup (h_{1}\circ f))(u)
\end{equation*}
\end{proof}

\section{Final remarks}

The intersection and the union of the systems are dual concepts and their
properties, as expressed by the previous theorems, are similar.

On the other hand, let us remark the roots of our interests in the Romanian
mathematical literature represented by the works in schemata with contacts
and relays from the 50's and the 60's of Grigore Moisil. Modeling is
different there, but the modelled switching phenomena are exactly the same
like ours.

\end{document}